\documentclass[aps,prc,twocolumn,nofootinbib]{revtex4-1}

\usepackage{graphicx}
\usepackage{epstopdf}
\usepackage{subfigure}
\usepackage{epsfig}

\usepackage{amsmath,amssymb,amsfonts}

\usepackage{color}

\begin{document}

\title{Thermalization of charm quarks in infinite and finite QGP matter}

\author{Shanshan Cao and Steffen A. Bass}
\affiliation{Department of Physics, Duke University, Durham, NC 27708, USA}

\date{\today}


\begin{abstract}

We study the thermalization process of charm quarks in hot and dense matter. 
The diffusion of heavy quarks is calculated via a Langevin equation, both for a static medium as well as for a QGP medium generated by a (3+1)D hydrodynamic model. 
We define two criteria for the thermalization of the heavy quarks, and
observe thermalization times that are longer than the lifetime of the QGP phase for reasonable values of the diffusion constant.

\end{abstract}
\maketitle


\section{Introduction}

The first decade of RHIC operations
have yielded a vast amount of interesting and
sometimes surprising data.
It is now generally accepted that RHIC has created
a hot and dense state of deconfined quark-gluon matter with properties similar to
that of an ideal fluid -- this state of matter
has been termed the {\em strongly interacting Quark-Gluon-Plasma} (sQGP) \cite{Gyulassy:2004zy}.
Among the unexpected observations to contribute to the moniker ``strongly interacting"
are the surprisingly large values of the elliptic flow coefficient $v_2$ and the
surprisingly small values of nuclear modification factor $R_{AA}$ \cite{Kelly:2004qw,Laue:2004tf,Adler:2005ab,Adler:2005xv, Abelev:2006db,Adare:2006nq,Adare:2010de} exhibited 
by D mesons, which are thought to be due to charm quarks interacting strongly with 
the QGP prior to hadronization \cite{Moore:2004tg,vanHees:2004gq}. 

In principle, the observed behavior of D mesons could be indicative of charm
quarks thermalizing in the QGP. It is therefore of great interest to study
the dynamics of heavy quarks in an expanding QGP and to verify whether they indeed approach equilibrium on a timescale on the order of the lifetime of the QGP. 
Due to the ``dead-cone effect" \cite{Dokshitzer:2001zm,Armesto:2005iq}, heavy quark interactions
with the QGP are dominated by quasi-elastic scattering off light quarks and gluons \cite{Mustafa:2003vh,Mustafa:2004dr}. Radiative contributions to e.g. the energy-loss
exist, but have been shown to be small (10-20\%) for heavy quarks at small and
intermediate transverse momenta \cite{Djordjevic:2007at}.
In the limit of multiple interactions, the heavy quark propagation through
a QGP can therefore be treated 
in the framework of a Langevin equation\footnote{Note that the Langevin equation may not be a good approximation if the momentum transfer during each interaction is large. See \cite{Moore:2004tg,Akamatsu:2008ge,Gossiaux:2008jv} for a discussion on this topic and other constraints
on the Langevin approach.} \cite{Svetitsky:1987gq,GolamMustafa:1997id,Moore:2004tg,Gossiaux:2009hr,Gossiaux:2011ea}.
In \cite{Moore:2004tg} it was estimated that the charm quark thermalization time is 
of the order of 7~fm/c. It was also suggested that the presence 
of resonant heavy-light quark interactions at moderate QGP temperatures accelerates 
the kinetic equilibration of charm quarks \cite{vanHees:2004gq,vanHees:2005wb} and reduces the 
thermalization time from 
over 30~fm/c (as estimated by previous pQCD calculations \cite{Svetitsky:1987gq,Combridge:1978kx}) to only a few fm/c.

In this paper we shall use a Langevin equation to conduct an 
investigation into heavy quark thermalization for infinite QGP matter at fixed temperature 
and then compare our  findings to a dynamical scenario in which the heavy 
quarks propagate through an expanding and cooling QGP medium, modeled with a (3+1)D 
relativistic hydrodynamic approach \cite{Nonaka:2006yn}.
Our goal is not to address current experimental data, but to answer the question
whether general features seen in the data, such as the presence of elliptic flow
and a small value of the nuclear modification factor, can be used to conclude that
heavy quarks actually have thermalized in the QGP medium created in ultra-relativistic
heavy-ion collisions or whether the heavy quarks remain off-equilibrium during their
entire evolution, despite exhibiting a strong response to the surrounding QGP.

This paper will be organized as follows. In the next section, we will discuss
the methodology used in our investigation by briefly summarizing the Langevin algorithm and introducing the criteria we use to test for heavy quark thermalization. In Sec.III, we will study the thermalization of charm quarks in a static QGP medium at fixed temperature 
and examine how the medium temperature, initial momentum of the charm quarks and the diffusion coefficient influence the thermalization process. In Sec.IV, the thermalization process in an
expanding and cooling QGP medium will be simulated. A summary and outlook will be provided at the end.

\section{Methodology}
\subsection{Heavy-quark dynamics}

Due to the ``dead-cone effect", heavy quark interactions
with the QGP are dominated by quasi-elastic scattering off light quarks and gluons.
In the limit of multiple interactions, the heavy quark propagation through
a QGP can therefore be treated 
in the framework of a Langevin equation \cite{Moore:2004tg}:
 
\begin{equation}
\frac{d\vec{p}}{dt}=-\eta_D(p)\vec{p}+\vec{\xi}.
\end{equation}
In principal, the noise term \begin{math} \vec{\xi} \end{math} may depend on the momentum of the particle, but we assume for simplicity that this is not the case  here. The random momentum kicks satisfy the following correlation relation:
\begin{equation}
\langle\xi^i(t)\xi^j(t')\rangle=\kappa\delta^{ij}\delta(t-t').
\end{equation}
\par
In order to simulate the momentum evolution, we adopt the Ito discretization:
\begin{equation}
\vec{p}(t+\Delta t)=\vec{p}(t)-\vec{d}_{Ito}(\vec{p}(t))\Delta t+\vec{\xi}\Delta t,
\end{equation}
\begin{equation}
\label{noise}
\langle\xi^i(t)\xi^j(t-n\Delta t)\rangle=\frac{\kappa}{\Delta t}\delta^{ij}\delta^{0n},
\end{equation}
where
\begin{equation}
\vec{d}_{Ito}(\vec{p})=\eta_D(p)\vec{p}.
\end{equation}
If the transfer of energy is small, it can be shown  that the fluctuation-dissipation relation applies, which indicates:
\begin{equation}
\eta_D(p)=\frac{\kappa}{2TE}.
\end{equation}
\par
Furthermore, according to Eq.(\ref{noise}), Gaussian noise with width 
\begin{math} \Gamma=\sqrt{\left. \kappa \right / \Delta t} \end{math}
will be used to generate the random momentum kicks in our calculation. The diffusion coefficient is related to the drag term via:
\begin{equation}
D=\frac{T}{M\eta_D(0)}=\frac{2T^2}{\kappa}.
\end{equation}

\subsection{QGP medium}

For our static infinite matter calculations, the only information on the 
medium that is required is the temperature, which remains fixed throughout
the time-evolution of the heavy quarks. For an expanding QGP,
we utilize a fully (3+1)D relativistic ideal hydrodynamic model \cite{Nonaka:2006yn}.
The initial conditions of the hydrodynamic calculation are tuned to describe the
hadronic data in the soft sector, 
such as hadron yields, spectra, rapidity-distributions as well 
as radial and elliptic flow. The hydrodynamic model provides us with the time-evolution
of the spatial distribution of temperature and collective flow velocity. 
Using our knowledge of the local flow velocity, for every Langevin time step we 
boost the heavy quark to the local rest frame
of the fluid cell through which it propagates. In the rest-frame of the cell
we perform the Langevin evolution (at the local temperature of the cell) before
boosting back to the global CM-frame.

Among the key assumptions of the hydrodynamic calculation is that the QCD medium experiences 
a sudden thermalization (to form a QGP) at an initial time $\tau_0$ (chosen here
to be 0.6~fm/c) at which the hydrodynamic evolution commences. The pre-equilibrium
phase of the medium cannot be modeled via hydrodynamics -- here we assume that
our heavy quarks stream freely prior to QGP formation. Any heavy quarks leaving
the QGP will stream freely as well. In our analysis we only treat deconfined
degrees of freedom and restrict our analysis to questions which can be addressed
in that context.

\subsection{Thermalization criterion}

Our thermalization criterion is based on the heavy quark
energy and its momentum components. For a medium at fixed temperature
without any inherent collective flow, an ensemble of thermalized
heavy quarks has the following energy distribution that allows for
the straightforward extraction of its ``temperature'' via an exponential fit:
\begin{equation}
\label{efit}
\frac{dN}{pEdE}=C e^{-E/T}.
\end{equation}
While this particular form for the energy distribution provides a convenient
representation for the extraction of the ``temperature'' of the heavy quark ensemble,
it is insufficient to actually indicate thermalization, since we still need to 
verify isotropy of particle momenta:
\begin{equation}
\label{}
f(p_i)\,=\, C \cdot T(\sqrt{p_i^2+m^2}+T)e^{-\sqrt{p_i^2+m^2}/T}.
\end{equation}
Note that if e.g. we initialize our heavy quark ensemble with a 
finite momentum along a given coordinate axis, its momentum 
distribution along that axis will be blue-shifted -- this can 
be taken into account by shifting the distribution along that axis using
a parameter \begin{math} \tilde{p_i} \end{math}; e.g. for the $z$ axis this 
would give 
\footnote{Rigorously, $p_z$ should be boosted via $\gamma p_z+\gamma \beta E$. However, it is found that in our study, for $\beta$ not too large (below 0.8), the much more convenient Eq.(\ref{pfit}) already fits the spectrum well (with an error less than 5\% for $T$).}:
\begin{equation}
\label{pfit}
f(p_z)=C \cdot T(\sqrt{(p_z-\tilde{p_z})^2+m^2}+T)e^{-\sqrt{(p_z-\tilde{p_z})^2+m^2}/T}.
\end{equation}

For an expanding and cooling QGP medium, establishing thermalization requires
the following procedure: for a given time step we select all cells in our
hydrodynamic medium within a temperature band of $T \pm \Delta T$. We then boost
all charm quarks localized in those cells into the respective local rest frames
of the cells they are residing in and 
calculate the resulting heavy quark energy and momentum distributions.
If the energy and momentum distributions of the heavy quark ensemble yield thermal
distributions with a temperature that lies within our selected temperature band,
we can conclude that the selected heavy quark ensemble has thermalized in the medium
at the given temperature and selected time step. This procedure can then be repeated
for other temperature bands and time steps.

\section{Equilibration in a Static Medium}
\subsection{Energy vs. momentum distributions as thermalization measures}

We start our investigation by 
simulating the diffusion of charm quarks in an infinite medium with a static temperature 
of 300~MeV. The initial momentum of the charm quarks is chosen to be 5~GeV/c in the \begin{math} \hat{z} \end{math} direction.  The diffusion coefficient is set to be 
\begin{math} D = 6/(2\pi T) \end{math}, which is the value that was found in 
\cite{Moore:2004tg} to  provide the best agreement to data on the elliptic flow and
the nuclear modification factor of non-photonic electrons \cite{Adare:2010de}. It is worth mentioning that different models for the medium may influence the value of the transport coefficients required to fit the experimental data \cite{Gossiaux:2011ea}.
\par

\begin{figure}[tb]
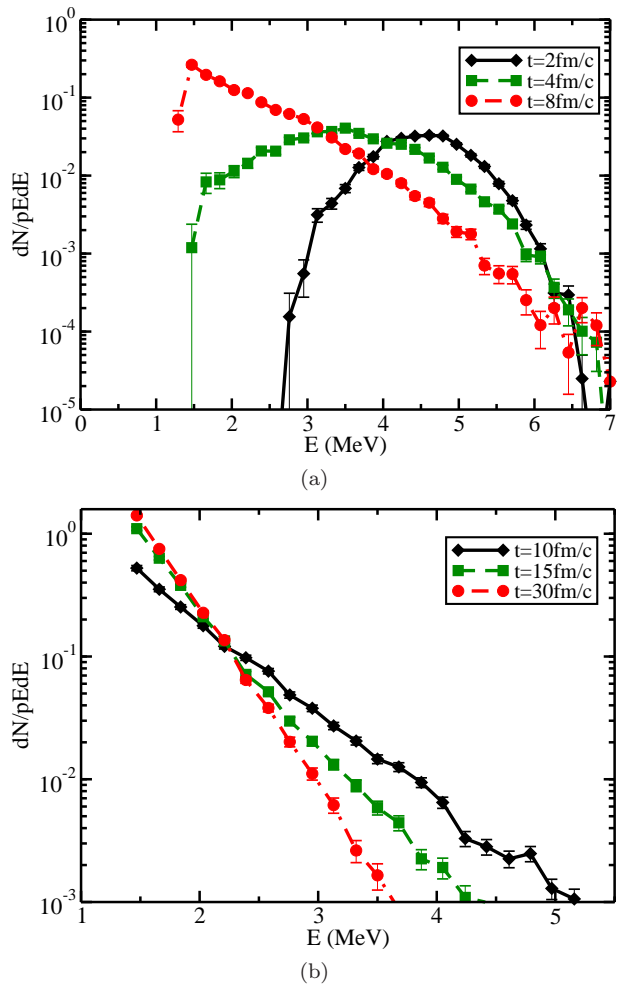

 \subfigure[]{\includegraphics[width=0.45\textwidth]{t2-8.eps}}
 \subfigure[]{\epsfig{file=t10-30.eps, width=0.45\textwidth, clip=}}
 \caption{(Color online) The evolution of the energy spectrum with respect to time. (a) shows the results between 2~fm/c and 8~fm/c, where no linear relation is observed; and (b) shows the results between 10~fm/c and 30~fm/c, where the linear relation is apparent.}
 \label{espectrum}
\end{figure}

Fig.\ref{espectrum} shows the energy spectrum $dN/pEdE$ vs. $E$ for different diffusion times. We see that a linear relation between $\ln(dN/pEdE)$ and $E$ does not occur for
diffusion times shorter than approx. 10~fm/c. The distribution appears thermal for longer
diffusion times. However, the slope continues to increase as a function of diffusion
time and does not converge to the temperature of the medium until a diffusion time of around 30~fm/c.
Therefore, despite  the shape of the energy distribution,
our ensemble of charm quarks is not fully equilibrated for diffusion times
between 10~fm/c and 30~fm/c. 
\par

\begin{figure}[tb]
 \includegraphics[width=0.45\textwidth]{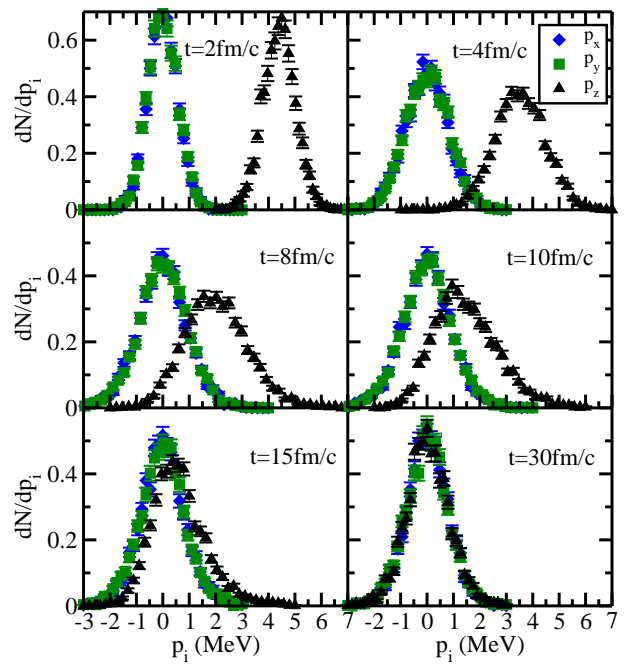}
 \caption{(Color online) The evolution of the momentum spectrum with respect to time.}
 \label{pspectrum}
\end{figure}

Fig.\ref{pspectrum} shows the momentum distributions in the three directions for the
same diffusion times as depicted in Fig.\ref{espectrum}. Since the initial momentum
of the charm quarks is in the \begin{math} \hat{z} \end{math} direction, the \begin{math} p_x \end{math} and \begin{math} p_y \end{math} spectra are symmetric with respect to  0 and are virtually identical to each other. The center of the $p_z$ spectrum, which initially
is a delta function at $p_z = 5$~GeV/c due to our initial condition,
shifts towards 0 as a function of increasing diffusion time,
signifying the influence of the drag term of the Langevin equation on the dominant
direction of propagation.
The widths of the momentum distributions along all three coordinate axes start to agree
with each other for diffusion times later than approx. 20~fm/c, hinting at a 
common ``temperature". At that time, full isotropy of the momenta is obtained.

\begin{figure}[tb]
 \epsfig{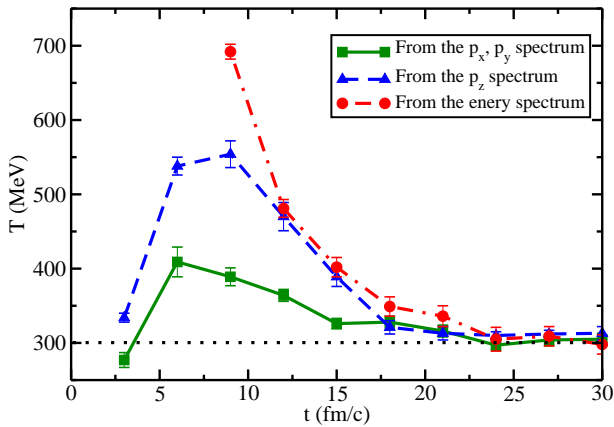}
 \caption{(Color online) A comparison of ``temperatures'' obtained from $p_x$, $p_y$, $p_z$ and the energy spectra. Since $\hat{x}$ and $\hat{y}$ directions are symmetric, we take the mean value of the ``temperatures'' fitted from $p_x$ and $p_y$ spectra here.}
 \label{Tcompare}
\end{figure}

We can obtain the ``temperature" evolution of the charm quark
ensemble by fitting the momentum distributions with Eq.(\ref{pfit}), and compare
the respective values of the temperature parameter with those obtained from the slope of the energy distribution. The results can be seen in Fig.\ref{Tcompare} as a function of diffusion
time, which helps to summarize our main observations from the previous figures
as follows:
\begin{itemize}
\item for diffusion times less than 10~fm/c, the \begin{math} p_z \end{math} spectrum is distinctly separated from the \begin{math} p_x \end{math} and \begin{math} p_y \end{math} spectra. The temperature parameters extracted from the widths of the distributions
initially rise and are significantly above the actual temperature of the medium. They
are of different values for the transverse momentum distributions  vs. the longitudinal
momentum distribution (as defined by the $z$ axis).
No temperature can be obtained from the energy distribution since no linear relation is observed at those short diffusion times.
\item for diffusion times between 10~fm/c and 20~fm/c, all momentum distributions
as well as the energy distribution exhibit a thermal shape, even though the extracted
temperature parameters strongly differ among each other and from the temperature
of the medium. Interestingly, the temperature parameter extracted from the longitudinal 
momentum distribution seems to track that from the energy distribution during this
time interval. However, both are significantly higher than the temperature parameter
extracted from the transverse momentum distributions.
\item at a diffusion time of roughly 25~fm/c, all temperature parameters agree
well with each other and have converged to the temperature of the medium, signaling
full equilibration of our charm quark ensemble.
\end{itemize}
Based on our observations, we shall 
define a ``quasi-equilibrium'' stage to be a near equilibrium state where the linear relation of the $\ln(dN/pEdE)$ vs. $E$ distribution appears and the temperature parameters extracted from energy and $p_z$ spectra are of approximately the same value. 
During this stage, our ensemble of charm quarks exhibits thermal properties,
even though it has not yet fully equilibrated with the surrounding medium. In contrast, a ``full equilibrium" is approached when the temperature parameters extracted from different ways agree with each other and reach the temperature of the medium.
\par

\subsection{The Blue Shift Effect}

To further understand the nature of the quasi-equilibrium state and why the ``temperature" of the charm quarks is still higher than that of the medium in this stage of the evolution, 
despite the linear relation between the $\ln(dN/pEdE)$ vs. $E$ distribution,  we may have to take into account a ``blue shift" caused by the center of mass motion of the charm quarks:
for our analysis, all charm quarks were initialized to carry a momentum of 5~GeV/c in the
$\hat z$ direction. This initial momentum can be seen as a center of mass motion -- it contributes
to the total energy of the particle, but is non-thermal in origin. As the charm quark
diffuses through the medium, this initial kinetic energy will dissipate through
the drag term of the Langevin equation. During this dissipation dominated phase
of the charm quark evolution, the center of mass motion will contribute to an additional part of energy and therefore a higher ``temperature''. In this sense, the quasi-equilibrium state can be understood as a stage when the thermal part of heavy quark motion is already close to equilibrium but the center of mass motion has not entirely dissipated.
\par
This blue shift can be verified by fitting Eq.(\ref{pfit}) with a momentum distribution 
that includes a momentum shift  \begin{math} \tilde{p_z} \end{math}. We will see that the blue shift is suppressed if the initial momentum of the heavy quarks is small enough that it becomes
comparable to its thermal motion. This in fact helps prove that the blue shift is the reason for a higher heavy quark ``temperature'' in the quasi-equilibrium state.

\subsection{Parameter Dependencies}

\begin{figure}[tb]
 \epsfig{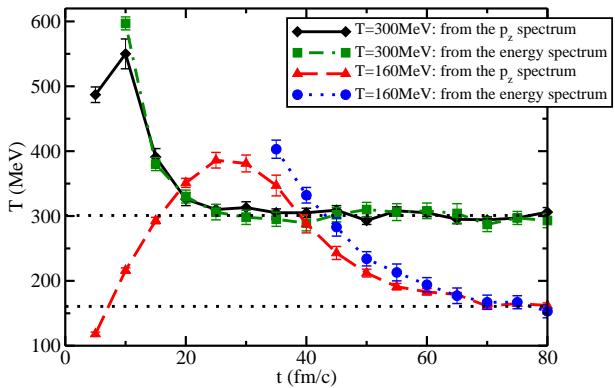}
 \caption{(Color online) A comparison of the charm quark thermalization processes in static media with different temperatures.}
 \label{Tinfluence}
\end{figure}

\begin{figure}[tb]
 \epsfig{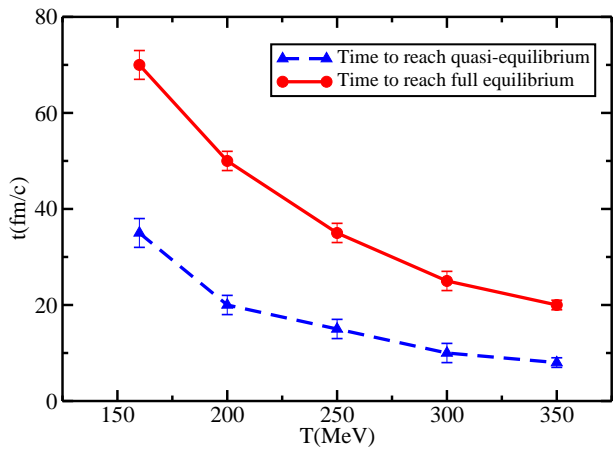}
 \caption{(Color online) The variation of critical times of thermalization with respect to the medium temperature.}
 \label{ctTdep}
\end{figure}

Let us now investigate the temperature dependence of the thermalization time:
as in the previous sections, we initialize charm quarks with 5~GeV/c momentum along the \begin{math} \hat{z} \end{math} direction. The diffusion coefficient is set to be \begin{math} D=6/(2\pi T) \end{math}. Here, we vary the temperature of the static medium and examine its corresponding influence on the thermalization process. The results are shown in Fig.\ref{Tinfluence} and Fig.\ref{ctTdep}.
\par

In Fig.\ref{Tinfluence}, we compare the  thermalization of charm quarks in a  300~MeV
temperature medium to that in a medium at 160~MeV. 
For charm quarks with an initial momentum of 5~GeV/c, their ``temperature" extracted
from the \begin{math} dN/dp_z\end{math}
distribution increases first, and then decreases until it approaches the medium temperature at which time the full equilibrium is reached. It is observed that the turning point between the rise and fall of the ``temperature" is close to the onset of the quasi-equilibrium stage. A similar trend can be observed via the
\begin{math} dN/dp_{x,y}\end{math}
distribution (not shown here), but the range of variation in the extracted ``temperature" is not as large as that from the 
\begin{math} dN/dp_z\end{math}
distribution. The above observation can also be obtained from media with other temperatures indicated in Fig.\ref{ctTdep} (not shown here).
\par

We can understand the ``turning point" behavior as follows. Since the charm quarks are all sampled with 5~GeV initial momentum in the $\hat{z}$ direction here, they are initially at
zero temperature with respect to their center of mass frame. During the diffusion process, the momenta of the charm quarks are randomized and thus their ``temperature" increases. When the random motion approaches equilibrium, i.e., the entrance of the quasi-equilibrium state, the charm quarks' ``temperature" might be higher than the medium temperature if the initial
center of mass momentum of the ensemble has not entirely been dissipated. After that, this temperature parameter gradually decreases towards the medium temperature with the dissipation of the center of mass momentum. Such a rise in the temperature parameter before reaching the quasi-equilibrium results from this particular initialization of charm quarks, and will not occur for more realistic scenarios where charm quarks are initialized in more realistic ways.\footnote{The temperature parameter before quasi-equilibrium is approached provides some insight into the dynamics of thermalization, namely the interplay of momentum broadening vs. energy dissipation through collisional energy loss. For a fixed initial momentum, this interplay is more obvious than compared to an initialization with charm quarks of different initial momenta in the later sections of our work. However, the values of the ``temperature'' obtained in this domain are not indicative of thermal behavior but just a measure of the broadening of the charm quark momentum distribution, since Eq.(\ref{pfit}) is only strictly valid near equilibrium.}

\par
As the temperature of the medium decreases, both the time needed to reach quasi-equilibrium and to obtain full equilibrium increase. The time to enter the quasi-equilibrium stage can vary from 8~fm/c (for 350~MeV medium) to 35~fm/c (for 160~MeV medium), and the time to approach the full equilibrium can vary from 20~fm/c (for 350~MeV medium) to 70~fm/c (for 160~MeV medium). This variation is clearly shown in Fig.\ref{ctTdep}. Note that since the calculation is carried out for static medium, critical behavior near $T_c\approx 160$~MeV is not expected here.
\par
\begin{figure}[tb]
 \epsfig{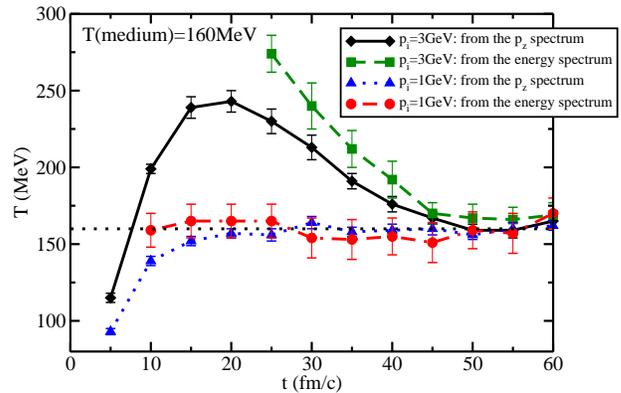}
 \caption{(Color online) A comparison of the thermalization processes of charm quarks with different magnitude of initial momenta in the $\hat{z}$ direction.}
 \label{pinfluence}
\end{figure}

In order to compare the results for the static medium with those for the QGP medium later, we need to simulate the diffusion of charm quarks with different initial momenta. Due to the expansion of the QGP medium itself, the initial momentum of the charm quarks in the local rest frame might be much smaller than that in the lab frame.  Therefore, in Fig.\ref{pinfluence}
we show our analysis for charm quarks with initial momenta of 1~GeV/c and 3~GeV/c in
a medium with 160~MeV temperature and compare 
to our previous results.
\par

We observe that the results of 3~GeV/c initial momentum are similar to those of the 5~GeV/c case shown in Fig.\ref{Tinfluence}: the ``temperature'' obtained from the
\begin{math} dN/dp_z\end{math}
distribution first increases and then decreases until the full equilibrium is approached, and the turning point corresponds to the entrance of the quasi-equilibrium stage. However, the results of the 1~GeV/c initial momentum case are different. There is no overshoot in the 
``temperature" for this case. Instead, the ``temperature'' obtained from the
\begin{math} dN/dp_z\end{math}
distribution keeps increasing gradually towards the temperature of the medium until the full equilibrium is reached. In other words, unlike in the higher initial momentum cases, the ``temperature" of the charm quarks with 1~GeV/c initial momentum is always below that of the medium until reaching full equilibrium. We attribute 
this difference to the suppression of the blue shift for the center of mass motion of the charm quarks, which is too small in this case to contribute significantly to the
energy of the charm quark. 
\par
For this low initial momentum situation, the entrance to the quasi-equilibrium stage can no longer be determined by the turning point, but only by the appearance of the linear relation in the 
$\ln (dN/pEdE)$ vs. $E$
distribution. In our simulation, for heavy quarks with 1~GeV/c initial momentum, this linear relation does not appear until 10~fm/c for the 160~MeV medium.
\par
Additionally, a comparison between Fig.\ref{Tinfluence} and Fig.\ref{pinfluence} suggests that with a decrease of the magnitude of the initial momentum, the thermalization occurs faster. For instance, the times needed for 5~GeV/c charm quarks to reach quasi-equilibrium and full equilibrium are 35~fm/c and 70~fm/c respectively, while these critical times for 1~GeV/c charm quarks are 10~fm/c and 20~fm/c. This decrease can also be understood with the concept of the blue shift: it takes less time for smaller initial momentum to dissipate.

\par

\begin{figure}[tb]
 \epsfig{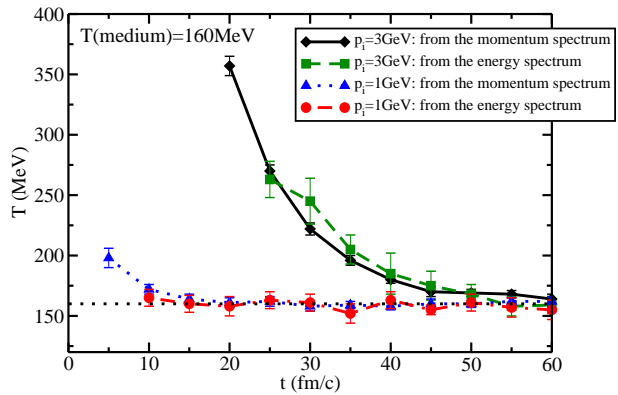}
 \caption{(Color online) The thermalization processes of charm quarks with initial momentum in random direction. Since $\hat{x}$, $\hat{y}$ and $\hat{z}$ directions are symmetric in this situation, we take the average of the ``temperatures'' obtained from $p_x$, $p_y$ and $p_z$ spectra.}
 \label{randomd}
\end{figure}

In order to verify the spatial invariance of our calculation, we randomize the direction of the initial momentum of the charm quarks and examine their thermalization process. The results are shown in Fig.\ref{randomd} for a medium temperature of 160~MeV. 
Comparing with Fig.\ref{pinfluence}, we observe that even though the trend of the evolution of the charm quark ``temperature'' may appear different, due to the coordinate axes now not reflecting anymore purely longitudinal or transverse motion respectively, the critical times for the charm quarks to enter the quasi-equilibrium state and the full equilibrium state are almost identical. For instance, both Fig.\ref{pinfluence} and Fig.\ref{randomd} indicate that for the 160~MeV static medium, charm quarks with 3~GeV/c initial longitudinal momentum enter the quasi-equilibrium stage at around 25~fm/c, and are entirely equilibrated at around 50~fm/c.  
Note that for an expanding QGP medium, the initial orientation of the motion of the 
charm quarks with respect the beam axis (i.e. the direction of longitudinal expansion)
will play a significant role.

\begin{figure}[tb]
 \epsfig{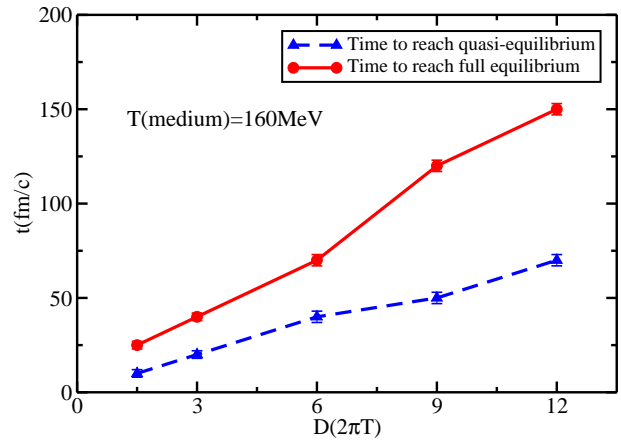}
 \caption{(Color online) The variation of critical times of thermalization with respect to the diffusion coefficient of the medium.}
 \label{ctDdep}
\end{figure}

To examine the influence of the diffusion coefficient on the process of heavy quark thermalization, we initialize the medium at a temperature of 300~MeV and the charm quarks
with an initial momentum of 5~GeV/c in the \begin{math} \hat{z} \end{math} direction, and vary the diffusion coefficient in accordance with the choices in \cite{Moore:2004tg}. The results are shown in Fig.\ref{ctDdep}.

The results indicate that as the diffusion coefficient decreases, or the drag coefficient increases, the thermalization process can be speeded up obviously. The times needed to enter the quasi-equilibrium state and the full equilibrium state can be respectively reduced from 70fm/c and 150fm/c for \begin{math} D=12/(2\pi T) \end{math} to 10~fm/c and 25~fm/c for \begin{math} D=1.5/(2\pi T) \end{math}.

\section{Analysis in a realistic QGP Medium}

In this section, we study the charm quark thermalization process in a realistic expanding QGP medium. The QGP medium is generated by a 3+1 dimensional ideal hydrodynamic calculation with initial conditions that have been adjusted to reproduce bulk properties of the QCD medium created in central Au+Au collisions at RHIC \cite{Nonaka:2006yn}.\footnote{Recent study \cite{Shen:2010uy} indicates the viscosity of the hydro medium may decrease the elliptic flow of $\pi^+$ and $p$, but does not have a significant influence on their spectra. Likewise, the temporal evolution of a viscous hydro (in terms of time-scale, flow velocities and temperatures) may differ at most by 20\% from the ideal hydrodynamic evolution used here -- this is well within the systematic parametric uncertainties of our study.} If not specified, the diffusion coefficient is chosen to be \begin{math} D=6/(2\pi T) \end{math} in this section following the choice most compatible with the data in \cite{Adare:2010de}. Although after 8~fm/c the medium temperature is below $T_c$ (160~MeV), we extend our study of charm quarks to the time of freeze-out (around 20~fm/c), assuming that D mesons interact with the medium in a similar way as the charm quarks do with QGP.
\par

\begin{figure}[tb]
 \epsfig{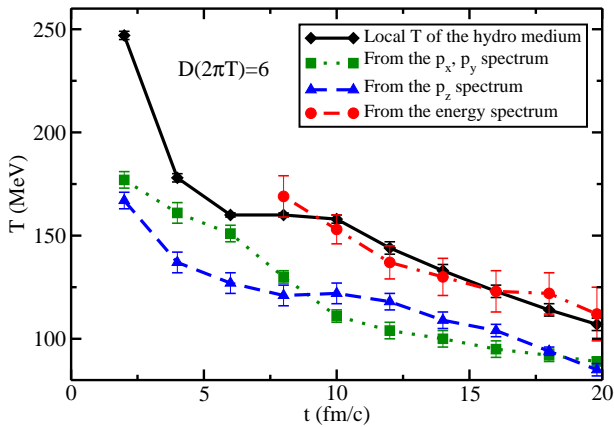}
 \caption{(Color online) The charm quark thermalization process in the QGP medium. The charm quarks have initial momentum of 5~GeV/c in the $\hat{z}$ direction.}
 \label{hydroz}
\end{figure}

\par
Let us first investigate the evolution of charm quarks 
 with 5~GeV/c initial momentum in the \begin{math} \hat{z} \end{math} direction. The results are
 shown in Fig.\ref{hydroz}. As can be seen, we are able to obtain the linear 
 relation from the 
$\ln (dN/pEdE)$ vs. $E$
distribution after 8~fm/c. Meanwhile, the ``temperature'' obtained from the 
\begin{math} dN/dp_z\end{math}
distribution is always below the local temperature of the QGP medium probed by the selected
sample of charm quarks. 
\par
By examining the value of $\tilde{p_z}$ in Eq.(\ref{pfit}), we find that the initial momentum of the charm quarks in the local rest frame is around 1.8~GeV/c. By comparing with the results of charm quarks with 1~GeV/c initial momentum in the 160~MeV static medium (shown in Fig.\ref{pinfluence}), we can conclude that the charm quarks here enter the quasi-equilibrium state after about 8~fm/c, but never reach the full equilibrium prior to freeze-out. In other words, the charm quarks are approaching the local temperature of the QGP medium but never catch up with it.
\par

\begin{figure}[tb]
 \epsfig{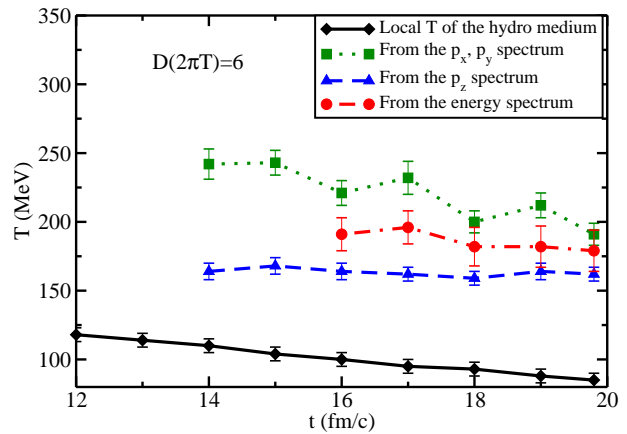}
 \caption{(Color online) The charm quark thermalization process in the QGP medium. The charm quarks have initial momentum of 5~GeV/c randomly distributed in the transverse plane. Since $\hat{x}$ and $\hat{y}$ directions are symmetric here, we take the mean value of the ``temperatures'' fitted from $p_x$ and $p_y$ spectra.}
 \label{hydrot}
\end{figure}

To bring the simulation closer to realistic conditions in  relativistic heavy-ion collisions, where charm quarks are distributed mostly throughout the mid-rapidity region, we set the initial momentum to be 0 in the \begin{math} \hat{z} \end{math} direction, and randomly distribute its direction in the transverse plane with a magnitude of 5~GeV/c. The results are shown in Fig.\ref{hydrot}. Unlike the static medium, the random distribution of the initial momentum direction increases the complexity of the thermalization process in the QGP medium. According to the occurrence of the linear relation of the $\ln (dN/pEdE)$ vs. $E$ distribution, the critical time to enter the quasi-equilibrium state is delayed to 16~fm/c and the charm quarks remain far off equilibrium at the time of freeze-out.
\par

\par
\begin{figure}[tb]
 \epsfig{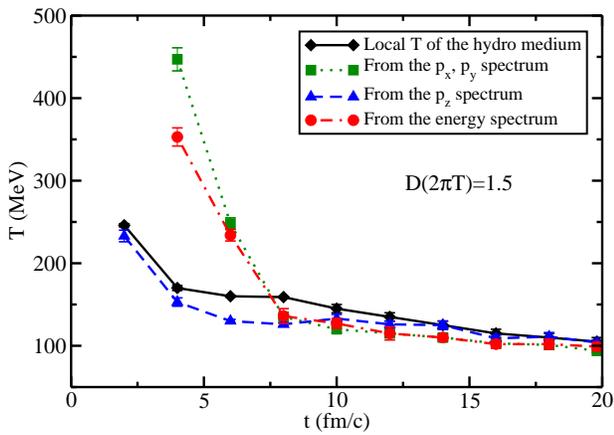}
 \caption{(Color online) The charm quark thermalization process in the QGP medium with small diffusion coefficient $D=1.5/(2\pi T)$. The charm quarks have initial momentum of 5~GeV/c randomly distributed in the transverse plane.}
 \label{hydrosmallD}
\end{figure}

If we want to see whether thermalization can be achieved in principle in an expanding
QGP medium, we need to increase the interaction strength of the charm quarks. This
can be realized by decreasing the diffusion coefficient to
 \begin{math} D=1.5/(2\pi T) \end{math}. For this value, we observe a significantly  faster thermalization. The results are shown in Fig.\ref{hydrosmallD}, which shows that the charm quarks enter the quasi-equilibrium stage as early as around 4~fm/c, and appear very close to full equilibrium  at the time of freeze-out.
\par

\begin{figure}[tb]
 \subfigure[]{\label{cascade6}\epsfig{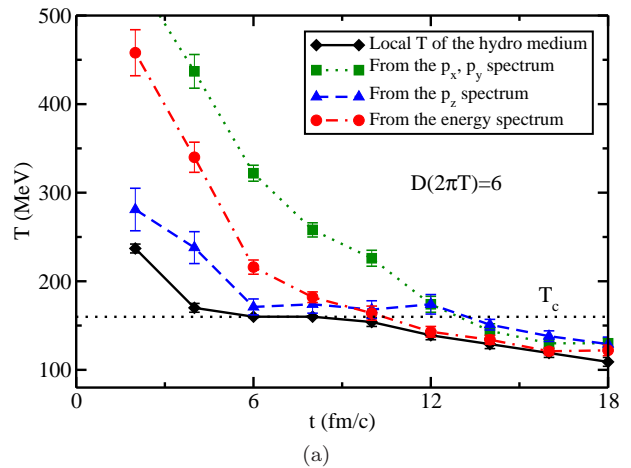}}
 \subfigure[]{\label{cascade1-5}\epsfig{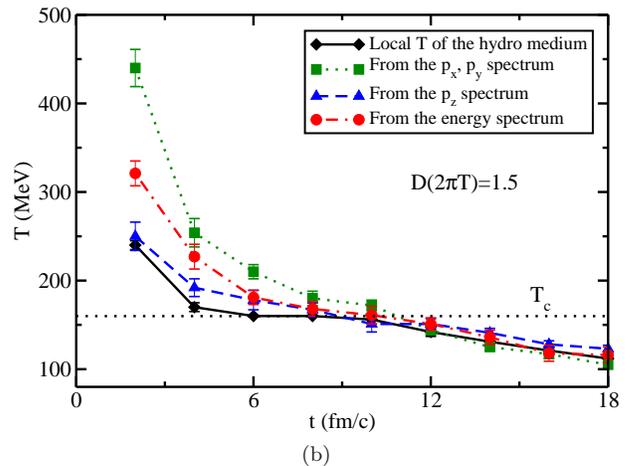}}
 \caption{(Color online) Thermalization of charm quarks with an initial distribution generated by the parton cascade model. (a) shows the results in the QGP medium with a diffusion coefficient of $D=6/(2\pi T)$; and (b) shows the results with a diffusion coefficient of $D=1.5/(2\pi T)$.}
 \label{cascade}
\end{figure}

Finally, we perform our analysis with charm quarks whose initial distribution is sampled from Au+Au collisions at RHIC energies generated by the VNI/BMS parton cascade model \cite{Geiger:1991nj, Bass:2002fh}. The medium evolution via a (3+1)D hydro and charm quark dynamics remain the same as in the
previous calculations. The c.m. energy is $\sqrt{s}=200$~GeV and the impact parameter is fixed at 2.4~fm for the generation of both the QGP medium and the initial charm quarks. The results are shown in Fig.\ref{cascade}.

As can be seen in Fig.\ref{cascade6}, with a diffusion coefficient of $D=6/(2\pi T)$, the ``temperature" of the charm quarks never manages to catch up with that of the medium until  freeze-out. A closer observation indicates that the charm quarks remain far off equilibrium during the entire lifetime of the QGP phase, i.e., when the medium temperature is above $T_c$. However, Fig.\ref{cascade1-5} suggests that if the diffusion coefficient is reduced to $D=1.5/(2\pi T)$, the thermalization process accelerates significantly and the ``temperature" of the charm quarks is able to catch up with that of the medium during its QGP phase, i.e., above $T_c$. Furthermore, a comparison between Fig.\ref{cascade} and Fig.\ref{hydrot}-\ref{hydrosmallD} indicates that charm quarks with an initial momentum distribution given by the cascade model thermalize faster than charm quarks with an initial momentum of fixed magnitude  of 5~GeV/c. This is due to the realistic PCM sample containing mostly charm quarks in the low momentum region, which are already rather close
to the thermal momentum scale.

\section{Summary and Outlook}
\par
In this paper, we have studied the dynamics of heavy quark thermalization in the framework of a Langevin equation. Simulations have been carried out in both static and dynamic QGP media, and our methodology allows for the extraction of the ``temperature" of the heavy quarks by fitting either the energy or the momentum spectrum. 

\par
Using an idealized static QGP medium, it is found that charm quark thermalization occurs
in two distinct steps: first a quasi-equilibrium is obtained in which the charm quark
energy distribution matches that of a thermal medium, but the momentum distribution
remains non-isotropic; subsequently the charm quark momenta isotropize and the charm quarks
are in full equilibrium with the surrounding medium.
The occurrence of this two step process might be explained by the blue shift effect due to the center of mass motion of the heavy quarks. We  define full equilibrium to imply that the ``temperature'' extracted from both methods of fitting the energy and the momentum distribution matches that of the medium. 
\par
Our simulations in the static medium indicates that as the medium temperature decreases, it takes longer time for the charm quarks to thermalize. Additionally, the thermalization speed is extremely sensitive to the diffusion coefficient of the medium: as the diffusion coefficient decreases, or the drag coefficient increases, the thermalization process can be significantly accelerated.  A decrease in the magnitude of the initial momentum leads to a suppression of the blue shift effect and therefore results in a faster thermalization process, due to the initial charm quark momenta being closer to the thermal momentum scale of the system.

\par
For a realistic expanding QGP medium, we find that for choices of the diffusion coefficient
that describe the elliptic flow and the nuclear modification factor of the charm quarks, 
thermalization does not occur within the lifetime of the QGP phase.
Only for a diffusion coefficient roughly $1/4$ of the desired value do the charm
quarks actually equilibrate. We find that the manifestation of collective behaviors, such as a significant elliptic flow, or the presence of a strongly interacting system (via the nuclear modification factor) is insufficient to conclude that charm quarks have actually thermalized in the medium. Our investigation indicates that their evolution occurs for the most part out of equilibrium, even though their properties are strongly
affected by the surrounding medium. It is worth emphasizing that this result does not imply a small value as $D=1.5/(2\pi T)$ for the diffusion coefficient, but only indicates that with a currently favored value of $D=6/(2\pi T)$, which describes experimental results well, charm quarks may remain off-equilibrium during their entire evolution although they exhibit a strong response to the surrounding QGP. 

The various parameter studies undertaken in our analysis should provide a measure for the  systematic uncertainties arising from the initial distribution of charm quarks and the variation of the diffusion coefficient. We conclude that the charm quarks' deviation from equilibrium during QGP's lifetime is significantly larger than one could overcome with different choices for the initial state in combination with realistic values for the diffusion coefficient around $D=6/(2\pi T)$. Of course, the microscopic structure of the charm quark scattering cross section in the non-perturbative regime remains largely unknown -- for example, it has been shown in \cite{vanHees:2004gq,vanHees:2005wb} that the introduction of resonant heavy-light quark interactions is able to speed up the heavy quark thermalization process. It has also been suggested that the memory effect due to strong correlations in the medium \cite{Asakawa:2006tc,Asakawa:2006jn,Asakawa:2007gf} may  accelerate the thermalization process. In addition, most current calculations do not account for any interaction in the pre-equilibrium phase of the medium prior to QGP 
formation. However, the strong chromo-electromagnetic field in this pre-equilibrium stage of QGP may cause a Weibel Instability \cite{PhysRevLett.2.83,Mrowczynski:1993qm,Romatschke:2003vc,Romatschke:2004au} and influence the heavy quark thermalization process. Furthermore, the recent study \cite{Abir:2011jb} indicates the radiation in the backward rapidity region may have a larger impact on the heavy quark energy loss than expected before.
\par
In a follow-up work, we shall address some of these open questions, in particular how
heavy quark thermalization is affected by memory effects and the pre-equilibrium phase of the QGP evolution.
\par

\begin{acknowledgements}
We are grateful to the Open Science Grid for providing computing
resources that were used in this work. We acknowledge many useful discussions
with Berndt M\"uller and Guangyou Qin and thank Christopher Coleman-Smith for
supplying us with charm quark distributions from the PCM.  This work was supported 
by U.S. department of Energy grant DE-FG02-05ER41367.

\end{acknowledgements}


\bibliographystyle{h-physrev5}
\bibliography{SABrefs}


\end{document}